\documentclass[comment, prd,twocolumn,nofootinbib, preprintnumbers, superscriptaddress]{revtex4}
\usepackage{amsmath,amssymb,bm,slashed,braket}
\usepackage{graphicx}
\usepackage{epstopdf}
\usepackage{float}
\usepackage{xcolor}
\usepackage{multirow}
\usepackage{dsfont}
\usepackage[colorlinks=true,
            linkcolor=blue,
            urlcolor=blue,
            citecolor=purple,          
            bookmarks=true,
            bookmarksnumbered=true,
            breaklinks=true,
            pdfpagemode=Fullscreen,
            pdfstartview=FitBH]{hyperref}
\usepackage[normalem]{ulem}
\allowdisplaybreaks[4]
\usepackage[capitalise]{cleveref}
\usepackage{orcidlink}

\usepackage{tikz} 
\usepackage{tkz-euclide}
\usetikzlibrary{backgrounds} 
\usetikzlibrary{decorations.markings}
\usetikzlibrary{decorations.pathmorphing}
\usetikzlibrary{arrows.meta}
\tikzset{
mystyle/.style={line width=1, baseline, scale=0.6, every node/.style={scale=1}},
v/.style={decorate, draw, decoration={snake, segment length=2.mm, amplitude=0.5mm}},
f/.style={draw, decoration={markings,mark=at position #1 with {\arrow[]{Latex[length=1.5mm,width=1.5mm]}}},
    postaction={decorate},node contents=#1},
f/.default=.6,
fb/.style={draw,decoration={markings,mark=at position #1 with {\arrowreversed[]{Latex[length=1.5mm,width=1.5mm]}}},
    postaction={decorate},node contents=#1},
fb/.default=.6,
s/.style={dashed,draw, decoration={markings,mark=at position #1 with {\arrow[]{Latex[length=1.5mm,width=1.5mm]}}},
    postaction={decorate},node contents=#1},
s/.default=.6,    
sb/.style={dashed,draw,decoration={markings,mark=at position #1 with {\arrowreversed[]{Latex[length=1.5mm,width=1.5mm]}}},
    postaction={decorate},node contents=#1},
sb/.default=.4,
snar/.style={dashed,draw,line width =1.25pt},
cross/.style={cross out, draw=black, minimum size=2*(#1-\pgflinewidth), inner sep=0pt, outer sep=0pt}, 
         }

\newcommand{\calO}{\mathcal{O}}
\newcommand{\C}{ {\tt C} }
\newcommand{\tL}{ {\tt L} }
\newcommand{\tR}{ {\tt R} }

\begin{document}

\title{Nucleon decays into one lepton plus two nonstrange mesons}

\author{Wei-Qi Fan}
\affiliation{State Key Laboratory of Nuclear Physics and
Technology, Institute of Quantum Matter, South China Normal
University, Guangzhou 510006, China}
\affiliation{Guangdong Basic Research Center of Excellence for
Structure and Fundamental Interactions of Matter, Guangdong
Provincial Key Laboratory of Nuclear Science, Guangzhou
510006, China}
\affiliation{School of Physics, Nankai University, Tianjin 300071, China}
\author{Yi Liao\,\orcidlink{0000-0002-1009-5483}}
\email{liaoy@m.scnu.edu.cn}
\affiliation{State Key Laboratory of Nuclear Physics and
Technology, Institute of Quantum Matter, South China Normal
University, Guangzhou 510006, China}
\affiliation{Guangdong Basic Research Center of Excellence for
Structure and Fundamental Interactions of Matter, Guangdong
Provincial Key Laboratory of Nuclear Science, Guangzhou
510006, China}
\author{Xiao-Dong Ma\,\orcidlink{0000-0001-7207-7793}}
\email{maxid@scnu.edu.cn}
\affiliation{State Key Laboratory of Nuclear Physics and
Technology, Institute of Quantum Matter, South China Normal
University, Guangzhou 510006, China}
\affiliation{Guangdong Basic Research Center of Excellence for
Structure and Fundamental Interactions of Matter, Guangdong
Provincial Key Laboratory of Nuclear Science, Guangzhou
510006, China}

\begin{abstract}
Nucleon decays into a lepton and two pseudoscalar mesons represent key channels for probing baryon number violation, complementing conventional two-body modes. 
In this paper, we model-independently correlate two- and three-body processes within the framework of low-energy effective field theory, 
performing a global analysis that avoids single-operator-dominance assumption. 
We derive significantly improved bounds on 15 three-body modes with a lepton ($e^+,\,\mu^+,\hat\nu=\nu/\bar\nu$) and two nonstrange mesons ($\pi,\eta$).
For charged-lepton modes,  our indirect lower limits on the partial lifetimes ($\Gamma^{-1}$) are more stringent than current Particle Data Group (PDG) values by more than three orders of magnitude.
For five (anti)neutrino modes, 
we establish for the first time $\Gamma^{-1}\gtrsim 10^{34}\,\rm yr$.  
Additionally, our analysis improves constraints on two-body processes $n\to e^+\pi^-$, $n\to \mu^+\pi^-$, and $p\to \hat\nu \pi^+$ by approximately a factor of 2 compared to the PDG limits. 
These results highlight the importance of leveraging correlations among different processes to better probe new physics, enabling more stringent constraints on experimentally challenging processes from well-measured ones. 
\end{abstract}

\maketitle

\textit{\textbf{Introduction}.} Nucleon decay searches are among the top physics goals of ongoing and upcoming neutrino experiments, including Super-Kamiokande (Super-K), JUNO~\cite{JUNO:2015zny}, Hyper-Kamiokande~\cite{Hyper-Kamiokande:2018ofw}, DUNE~\cite{DUNE:2020ypp}, and Theia~\cite{Theia:2019non}.
They stand out as the most practical avenues to test baryon number violation, which is predicted in various scenarios beyond the standard model, such as grand unified theories and simplified models involving leptoquarks.  

The conventional two-body decay modes involving a lepton and a pseudoscalar meson are prominent channels for testing BNV. 
Such modes are widely predicted across various theoretical models, 
and have been extensively searched for over the years, with experimental lower bounds on their partial lifetimes (i.e., the inverse decay width $\Gamma^{-1}$, which experimental papers typically express as $\tau/B$) reaching $\calO(10^{31-34})\,\rm yr$~\cite{ParticleDataGroup:2024cfk}. 
Theoretically, the same underlying interactions responsible for the two-body decay modes can also induce three-body nucleon decays involving a lepton and two pseudoscalar mesons ($\pi\pi$, $\pi\eta$, and $\pi K$).
In the latter case, the decay rates could be competitive with those of two-body modes~\cite{Wise:1980ch,Oset:1987tm}, since the additional meson is produced via strong interactions.
However, these three-body modes have received less experimental attention, and only the IMB-3 experiment conducted a preliminary search for several such channels over 27 years ago~\cite{McGrew:1999nd}, with the resulting bounds being about one to two orders of magnitude less stringent than those of two-body modes. 

The extraction of constraints on three-body decay modes from experimental data typically assumes a uniform phase space distribution and applies specific kinematic cuts to derive bounds. 
It also suffers from sizable uncertainties due to complex nuclear effects (such as Fermi motion, nuclear binding energy, nucleon correlations) and final-state interactions between mesons and nucleons.
As a result, the resulting bounds are less robust than those for two-body ones. 
Therefore, it is desirable to obtain independent bounds on three-body decay modes without relying on these theoretical assumptions.

In this paper,  
we use a model-independent effective field theory (EFT) approach to correlate two- and three-body nucleon decay modes. This allows us to derive improved bounds on three-body decay modes from existing, experimentally well-constrained limits on two-body modes. 
We focus on nonstrange decay channels, including 9 two-body modes ${\tt N}\to l\,M_1$ and 15 three-body modes  ${\tt N}\to l\,M_1\,M_2$, 
where ${\tt N}\equiv p,n$, $l\equiv e^+,\,\mu^+,\,\hat\nu$, and 
$M_{1,2}\equiv \pi^\pm,\,\pi^0,\,\eta$.
Here, $\hat\nu$ denotes either an antineutrino $\bar\nu$ or a neutrino $\nu$, with no distinction made among different flavors.
We take the leading-order dimension-6 (dim-6) BNV interactions in low-energy EFT (LEFT)~\cite{Jenkins:2017jig,Liao:2020zyx} as our starting point.
Using chiral perturbation theory (ChPT),  
we calculate the decay widths for all 24 modes 
as functions of the LEFT Wilson coefficients (WCs).

In contrast to the traditional analyses that assume a single-operator or single-parameter dominance, i.e., turning on one operator at a time, we perform a global analysis by simultaneously varying all relevant WCs for each decay mode.  
Using the well-constrained experimental bounds on 6 two-body processes [$p\to\ell^+(\pi^0,\eta)$ with $\ell=e,\mu$ and $n\to\hat{\nu}(\pi^0,\eta)$] as inputs, we constrain the multidimensional WC parameter space into a closed region for each field configuration. 
From the surviving multidimensional WC region in each case, we establish conservative new bounds on all remaining 18 modes, which include 3 two-body modes ($n\to\ell^+\pi^-$ and $p\to \hat\nu \pi^+$). 
For the 10 charged-lepton modes $p\to\ell^+(\pi^+\pi^-,\pi^0\pi^0,\pi^0\eta)$ and $n\to\ell^+\pi^-(\pi^0,\eta)$, the partial lifetimes are constrained to be of order $\calO(10^{35-36})$ yr.
For the 5 neutrino/antineutrino modes $p\to\hat\nu\pi^+(\pi^0,\eta)$ and $n\to\hat\nu(\pi^+\pi^-,\pi^0\pi^0,\pi^0\eta)$, the partial lifetime bounds are around $\calO(10^{34})$ yr. 

\begin{figure}[t]
\centering
\includegraphics[width=0.49\textwidth]{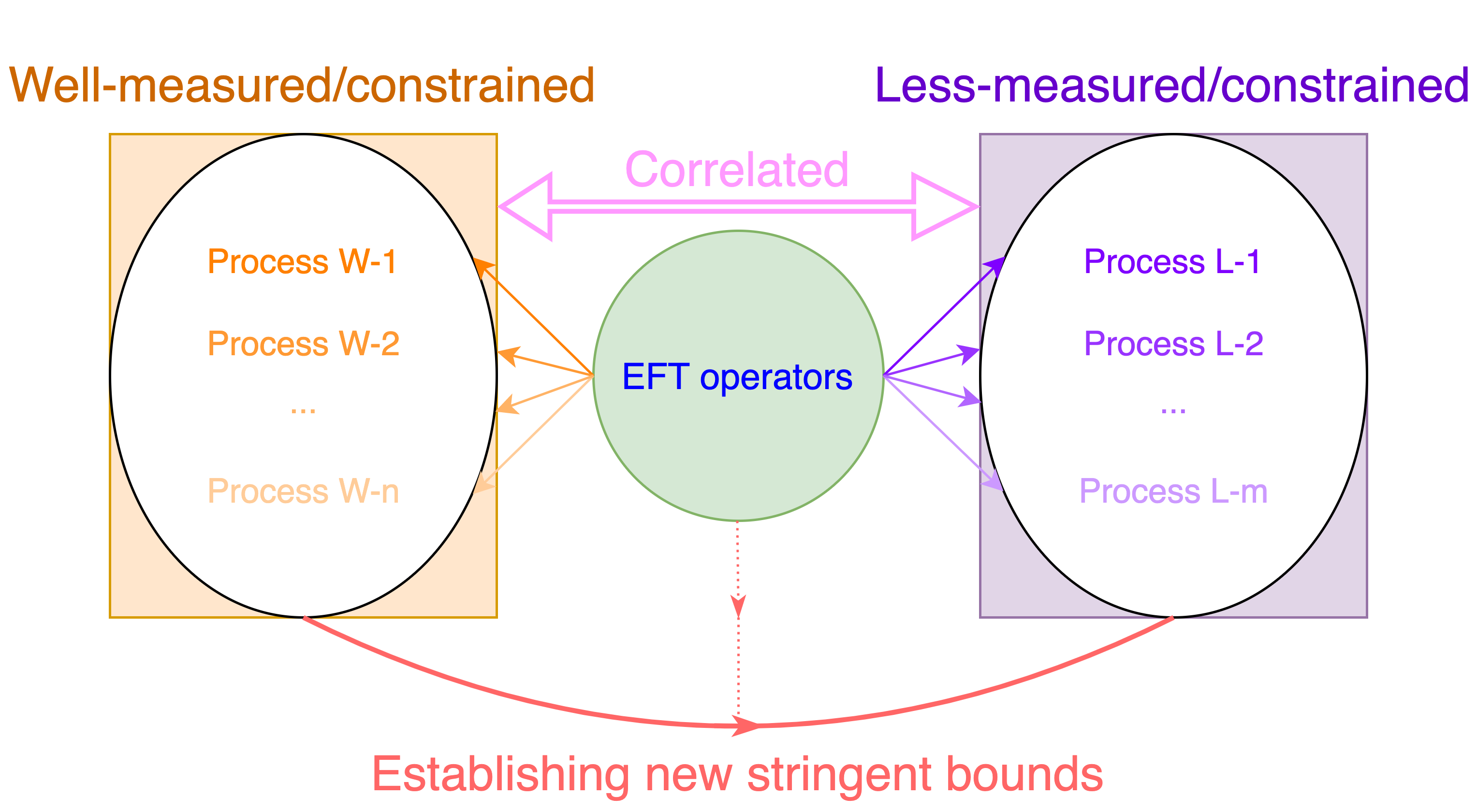}
\caption{Correlation diagram of different processes induced by a common set of EFT operators.}
\label{fig:correlation}    
\end{figure}

\textit{\textbf{Method}.} We explore correlations among different processes that are dominantly induced by a common effective Lagrangian composed of a shared set of effective operators.
Such correlations allow us to place more stringent constraints on less constrained processes by leveraging those that are already tightly constrained, as pictorially illustrated in \cref{fig:correlation}. 
This strategy underpins the SM: its parameters are extracted with high precision from well-measured processes and then used as inputs to predict other, less- or even not yet measured, observables. 
This approach is closely related to the global fit method; however, instead of using all available data, we select a small set of experimentally well-constrained observables.
From these, we derive closed-region bounds on the relevant WCs with their magnitudes being bounded from above. 
Using these bounded WCs, we then compute  limits for processes that are experimentally less constrained but governed by the same set of WCs.   

\textit{\textbf{EFT descriptions}.} Nucleon decays occur at an energy scale of $\calO(\rm GeV)$,
whereas the underlying BNV mechanism presumably lies at a scale that is many orders of magnitude higher. This energy hierarchy makes the LEFT a powerful framework for model-independent  parametrization of various scenarios using a small set of higher-dimensional operators. The effects of these operators diminish rapidly as their dimension increases. 
In particular, the leading-order LEFT operators that mediate two- and three-body nucleon decays involving a lepton and one or two nonstrange-quark mesons arise at dimension 6 and take the form 
\begin{subequations}
\label{eq:dim6Ope_lqqq}
\begin{align}
\calO_{\ell uud}^{\tL\tR}=(\overline{\ell_{\tL}^{\C}} u_{\tL}^\alpha) 
(\overline{u_{\tR}^{\beta \C}} d_{\tR}^\gamma)
\epsilon_{\alpha \beta \gamma},
\\
\calO_{\ell uud}^{\tL\tL}=(\overline{\ell_{\tL}^{\C}} u_{\tL}^\alpha) 
(\overline{u_{\tL}^{\beta \C}} d_{\tL}^\gamma)
\epsilon_{\alpha \beta \gamma},
\\
\calO_{\nu dud}^{\tL\tR}=(\overline{\nu_{\tL}^{\C}} d_{\tL}^\alpha) 
(\overline{u_{\tR}^{\beta \C}} d_{\tR}^\gamma)
\epsilon_{\alpha \beta \gamma},
\\
\calO_{\nu dud}^{\tL\tL}=(\overline{\nu_{\tL}^{\C}} d_{\tL}^\alpha) 
(\overline{u_{\tL}^{\beta \C}} d_{\tL}^\gamma)
\epsilon_{\alpha \beta \gamma},
\end{align}
\end{subequations}
plus their chirality partners with the interchange of the chiral projectors $\tL\leftrightarrow \tR$ and $\nu_\tL^\C\leftrightarrow\nu_\tL$.
Here, $\alpha,\beta,\gamma$ are color indices and $\C$ denotes charge conjugation.
Neglecting the tiny neutrino mass, the operators $\calO_{\nu dud}^{\tL\tR(\tL\tL)}$ and  
$\calO_{\bar\nu dud}^{\tR\tL(\tR\tR)}$ contribute exclusively to the antineutrino and neutrino modes, respectively. 
For later use, the corresponding WC of each operator $\calO_i$ is denoted by $C_i$. 
For instance, $C_{\ell udd}^{\tL\tR}$ corresponds to $\calO_{\ell udd}^{\tL\tR}$.

The four operators $\calO_{\ell uud}^{\tt XY}$ ($\tt X,Y=\tL,\tR$)
involving a charged lepton $\ell=e,\mu$ can generate 6 two-body modes $p\to\ell^+(\pi^0,\eta)$ and $n\to\ell^+\pi^-$, 
and 10 three-body modes $p\to\ell^+(\pi^+\pi^-,\pi^0\pi^0,\pi^0\eta)$ and 
$n\to\ell^+\pi^-(\pi^0,\eta)$. 
The other four operators involving a neutrino or an antineutrino field contribute to 3 two-body modes $n\to\hat\nu(\pi^0,\eta)$ and $p\to \hat\nu\pi^+$, 
and 5 three-body modes
$p\to\hat\nu\pi^+(\pi^0,\eta)$, $n\to\hat\nu(\pi^+\pi^-,\pi^0\pi^0)$, 
and $n\to \hat\nu \pi^0\eta$.
In this paper, we omit operators involving strange quarks and the corresponding nucleon decays into a kaon. 
For these cases,
the limited number of experimentally accessible modes does not suffice to constrain the relevant dim-6 WCs to a bounded region, thus precluding assumption-free bounds on the associated three-body modes. 
We postpone the single-operator-dominance analysis for these cases in \cite{Fan:2026csl}.

To calculate the hadronic matrix elements, we employ ChPT to match the LEFT interactions onto the hadronic counterparts, with the quark degrees of freedom replaced by those of the octet baryons and pseudoscalar mesons. The leading-order chiral Lagrangian takes the form~\cite{Claudson:1981gh,Liao:2025vlj}
\begin{align}
{\cal L}_{\slashed{B}}^{\tt ChPT} &=
c_1 {\rm Tr}\left[ 
{\cal P}_{  \bar{\pmb{3}}_{\tL} \otimes \pmb{3}_{\tR}} \xi B_{\tL} \xi 
- {\cal P}_{\pmb{3}_{\tL} \otimes \bar{\pmb{3}}_{\tR}} \xi^\dagger B_{\tR} \xi^\dagger  \right]
\nonumber
 \\
&+ c_2 {\rm Tr}\left[ 
{\cal P}_{\pmb{8}_{\tL} \otimes \pmb{1}_{\tR}} \xi B_{\tL} \xi^\dagger
- {\cal P}_{ \pmb{1}_{\tL} \otimes  \pmb{8}_{\tR}} \xi^\dagger B_{\tR} \xi
\right]+\text{H.c.},
\label{eq:BNVchiral}
\end{align}
where $\xi$ is the square root of the pseudoscalar meson matrix $\Sigma = \xi^2 ={\rm exp}[{2 i \Pi/f_{\pi}}]$, 
with $\Pi$ being the octet pseudoscalar meson fields and $f_{\pi}$ the pion decay constant in the chiral limit. 
$B_{\tL,\tR}$ are the left- and right-handed chiral octet baryon fields. 
${\cal P}_{i_\tL\otimes j_\tR}$ are spurion field matrices consisting of the products of LEFT WCs and non-QCD lepton fields of the LEFT operators, transforming as chiral irreducible representation $i_\tL\otimes j_\tR$ under the QCD chiral group $\rm SU(3)_\tL\otimes SU(3)_\tR$. 
Their specific forms related to the operators in \cref{eq:dim6Ope_lqqq} are provided in Eq.\,(5) of Ref.\,\cite{Fan:2024gzc}. 

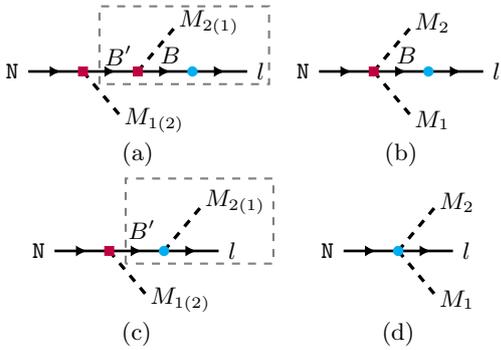
\begin{figure}[t]
\centering
\begin{tikzpicture}[mystyle,scale=0.8]
\begin{scope}[shift={(1,1.2)}]
\draw[thick,dashed,gray] (1.95,-0.35) rectangle (6.6,1.8);
\draw[f] (0, 0)node[left]{$\texttt{N}$} -- (1.5,0);
\draw[f] (1.5, 0) -- (3,0) node[midway, xshift=3pt, yshift = 6 pt]{\small$B'$};
\draw[f] (3,0) -- (4.5,0) node[midway,xshift = 2pt, yshift =  6 pt]{\small$B$};
\draw[snar, black] (1.5,0) -- (2.5,-1.2) node[right,xshift = -2pt, yshift = -2 pt]{$M_{1(2)}$};
\draw[f] (4.5, 0) -- (6,0) node[right]{$l$};
\draw[snar, black] (3,0) -- (4,1.2) node[right,xshift = -2pt, yshift = 2 pt]{$M_{2(1)}$};
\filldraw [purple] (1.4,-0.1) rectangle(1.6,0.1);
\filldraw [purple] (2.9,-0.1) rectangle(3.1,0.1);
\filldraw [cyan] (4.5,0) circle (3pt);
\node at (3,-2.3) {(a)};
\end{scope}
\end{tikzpicture}
\hspace{0.1cm}%
\begin{tikzpicture}[mystyle,scale=0.8]
\begin{scope}[shift={(1,1.2)}] 
\draw[f] (0, 0)node[left]{$\texttt{N}$} -- (1.5,0);
\draw[f] (1.5, 0) -- (3,0) node[midway,xshift = 2pt, yshift = 6 pt]{\small$B$};
\draw[snar, black] (1.5,0) -- (2.5,-1.2) node[right,xshift = -2pt, yshift = -2 pt]{$M_1$};
\draw[f] (3.0, 0) -- (4.5,0) node[right]{$l$};
\draw[snar, black] (1.5,0) -- (2.5,1.2) node[right,xshift = -2pt, yshift = 2 pt]{$M_2$};
\filldraw [purple] (1.4,-0.1) rectangle(1.6,0.1);
\filldraw [cyan] (3,0) circle (3pt);
\node at (2.25,-2.3) {(b)};
\end{scope}
\end{tikzpicture}
\\%
\begin{tikzpicture}[mystyle,scale=0.8]
\begin{scope}[shift={(1,1.2)}] 
\draw[thick,dashed,gray] (1.95,-0.35) rectangle (6,2);
\draw[f] (0, 0)node[left]{$\texttt{N}$} -- (1.5,0);
\draw[f] (1.5, 0) -- (3,0) node[midway,xshift=2pt, yshift = 8 pt]{\small$B'$};
\draw[snar, black] (1.5,0) -- (2.5,-1.2) node[right,xshift = -2pt, yshift = -2 pt]{$M_{1(2)}$};
\draw[f] (3.0, 0) -- (4.5,0) node[right]{$l$};
\draw[snar, black] (3,0) -- (4,1.2) node[right,xshift = -2pt, yshift = 2 pt]{$M_{2(1)}$};
\filldraw [purple] (1.4,-0.1) rectangle(1.6,0.1);
\filldraw [cyan] (3,0) circle (3pt);
\node at (2.25,-2.3) {(c)};
\end{scope}
\end{tikzpicture}
\hspace{0.4cm}%
\begin{tikzpicture}[mystyle,scale=0.8]
\begin{scope}[shift={(1,1.2)}] 
\draw[f] (0, 0)node[left]{$\texttt{N}$} -- (1.5,0);
\draw[f] (1.5, 0) -- (3,0) node[right]{$l$};
\draw[snar, black] (1.5,0) -- (2.5,1.2) node[right,xshift = -2pt, yshift = 2 pt]{$M_2$};
\draw[snar, black] (1.5,0) -- (2.5,-1.2) node[right,xshift = -2pt, yshift = -2 pt]{$M_1$};
\filldraw [cyan] (1.5,0) circle (3pt);
\node at (1.5,-2.3) {(d)};
\end{scope}
\end{tikzpicture}
\caption{Leading-order Feynman diagrams contributing to three-body nucleon decays that produce one lepton $l=e^+,\,\mu^+,\,\hat\nu$ and two pseudoscalar mesons $M_{1,2}=\pi,\eta$.
The intermediate baryon states $B,\,B'=p,\,n$. 
The cyan blobs and purple squares represent the dim-6 BNV vertices and the SM strong interaction vertices, respectively. The subdiagrams enclosed by dashed boxes correspond to two-body nucleon  decays.}
\label{fig:Feyndiagram}
\end{figure}

Expanding the pseudoscalar meson matrix $\xi$ up to the quadratic order in meson fields, \cref{eq:BNVchiral} results in local baryon-lepton mixing terms ${\cal L}_{Bl}$, three-point baryon-lepton-meson vertices ${\cal L}_{BlM}$, and four-point vertices involving two mesons ${\cal L}_{BlM_1M_2}$. 
Together with the conventional baryon-meson vertices (${\cal L}_{BBM}$ and ${\cal L}_{BBMM}$) due to QCD strong interactions, we can form four leading-order Feynman diagrams contributing to a three-body nucleon decay process ${\tt N}\to l M_1 M_2$, as shown in \cref{fig:Feyndiagram}. 
The two-body decay modes correspond to the subdiagrams enclosed by dashed boxes, all of which share the same BNV vertices with the three-body decays. 
This implies that the same set of BNV interactions contributes to both processes. 
By calculating these diagrams and incorporating the numerical values of the relevant hadronic parameters, the decay widths can be expressed solely in terms of the LEFT WCs. 
The complete expressions for all 24 modes are collected in the Appendix,
and further calculational details will be given in our upcoming long paper \cite{Fan:2026csl}.
The results show that 
the 8 processes involving a charged lepton $l=e^+,\,\mu^+$ are described by the four WCs $C_{\ell uud}^{{\tt LR/RL/LL/RR}}$, 
those with an antineutrino $\bar\nu$ are described by the two WCs $C_{\nu dud}^{{\tt LR/LL}}$, 
and those with a neutrino $\nu$
are described by the two WCs $C_{\bar\nu dud}^{\tt RL/RR}$. 
For each set of processes, 
their decay widths share the same structure but differ only in the numerical prefactors multiplying the quadratic WCs.  

In the calculation, we have neglected  contributions mediated by vector mesons. 
Within resonance ChPT, 
the octet vector mesons can  
couple both to a pair of pseudoscalar mesons~\cite{Ecker:1989yg} and to a pair of baryons~\cite{Unal:2015hea}.
Combining with the BNV baryon-lepton mixing vertices in ${\cal L}_{Bl}$, the $\rho$ mesons can generate additional contributions to the three-body modes involving $\pi^+\pi^-$ or $\pi^\pm\pi^0$.
However, the numerical analysis of~\cite{Fan:2026csl} shows that such contributions are subdominant, amounting to less than 30\,\% of those from \cref{fig:Feyndiagram}.
We therefore neglect them. 
Furthermore, we have omitted contributions arising from the BNV baryon-lepton-vector-meson interactions in the local chiral Lagrangian involving vector mesons~\cite{Liao:2025sqt}, due to unknown values of the relevant low energy constants. It is reasonable to expect that these contributions are at most comparable to the ones we estimated above.

Examining the prefactors for each LEFT WC in the decay width expressions, 
our EFT calculation indicates that the decay rates for three-body modes involving two $\pi$s are suppressed by approximately a factor of 5-100 relative to those for two-body modes involving a single $\pi$, depending on which specific modes are compared. 
A similar suppression holds for modes involving an $\eta$ meson. 
In particular, for proton decays into a positron and one or two pions, we obtain the approximate relations   
$\Gamma(p\to e^+\pi^+\pi^-)\approx
0.13\,\Gamma(p\to e^+\pi^0)$
and 
$\Gamma(p\to e^+\pi^0\pi^0)\approx 0.021\,\Gamma(p\to e^+\pi^0)$. 
Despite the enhancement of the three-body decay amplitudes due to strong interactions, the phase space reduction factor $1/(8\pi)\approx0.04$ results in an overall suppression.   
Our above ratios differ from the final estimates in~\cite{Wise:1980ch} by approximately one order of magnitude. This discrepancy is a joint effect of presenting decay widths in terms of pion-pair isospin vs mass eigenstates and an oversimplified treatment of kinematic effects in their analysis, particularly the neglect of pion masses. 

While contributions from dim-6 LEFT operators to two-meson modes are typically suppressed compared to their contributions to single-meson modes, certain higher-dimensional operators may predominantly mediate two-meson processes. 
For instance, isospin analysis reveals that dim-9 $\Delta I=5/2$ operators can exclusively induce two-pion modes while suppressing single-meson modes. 
This happens because the one-meson modes ${\tt N}\to l (\pi,\eta)$ are restricted to isopsin changes of $\Delta I = 1/2$ or $3/2$, whereas the two-pion modes ${\tt N}\to l \pi\pi$ also allow $\Delta I =5/2$ transitions. 
However, realizing such scenarios would require significant cancellation or fine-tuning of the WCs.

\textit{\textbf{Analysis}.} With the expressions for the decay widths, we now establish constraints on each set of WCs that share the same field configuration but differ in chiralities using the associated two-body decay modes. 
For each such set, our numerical results indicate that only two two-body modes---one involving a $\pi$ and the other an $\eta$---are sufficient to constrain the multidimensional parameter space into a closed region, 
as they depend on the WCs in very different numerical combinations. 

To maximize the constraining power, we employ the currently most stringent Super-K 90\,\% CL bounds on 4 proton decay modes into a charged lepton~\cite{Super-Kamiokande:2024qbv,Super-Kamiokande:2020wjk}
\begin{subequations}
\label{eq:exp_bound_l}
\begin{align}
\Gamma^{-1}_{\tt SK}(p\to e^+\pi^0)
\gtrsim&\, 2.4\times 10^{34}\,\rm yr,
\\
\Gamma^{-1}_{\tt SK}(p\to e^+\eta)
\gtrsim&\, 1.4\times 10^{34}\,\rm yr,
\\
\Gamma^{-1}_{\tt SK}(p\to \mu^+\pi^0)
\gtrsim&\, 1.6\times 10^{34}\,\rm yr,
\\
\Gamma^{-1}_{\tt SK}(p\to \mu^+\eta)
\gtrsim&\, 7.3\times 10^{33}\,\rm yr.
\end{align}    
\end{subequations}
For the (anti)neutrino modes, we take the recent Super-K 90\,\% CL result~\cite{Super-Kamiokande:2025lxa} and the older IMB-3 90\,\% CL result~\cite{McGrew:1999nd} as our inputs:  
\begin{subequations}
\label{eq:exp_bound_nu}
\begin{align}
\Gamma^{-1}_{\tt SK}(n\to \hat\nu\pi^0)
\gtrsim&\, 1.4\times 10^{33}\,\rm yr,
\\
\Gamma^{-1}_{\tt IMB-3}(n\to \hat\nu\eta)
\gtrsim&\, 1.58\times 10^{32}\,\rm yr.
\end{align}    
\end{subequations} 
By parametrizing each WC in terms of its magnitude and phase, $C_i \equiv |C_i| e^{i \theta_i}$, we see that the charged-lepton modes involve seven independent parameters---four magnitudes and three phase differences, whereas the neutrino or antineutrino modes involve two magnitudes and one phase difference. 

To derive global constraints on these parameters associated with each field configuration, we randomly generate $10^7$ points in the multidimensional space, requiring that the phases lie in the range $[0,\,2\pi]$ and that the WC magnitudes range from zero to some chosen upper values.  
We then retain only those points that satisfy the condition that the theoretically computed partial lifetimes for each pair of modes are above the experimental 90\,\% lower limits given in \cref{eq:exp_bound_l,eq:exp_bound_nu}.
We conclude that the constraints form closed regions if the WC magnitudes are significantly smaller than the initially chosen upper values. 
In Ref.\,\cite{Fan:2026csl}, we will present all projected 2D planes spanned by the WC magnitudes and phases for the surviving points; and the bounded behavior can be clearly observed from them.    
In the following, we take the antineutrino modes as an example for illustration. 

\begin{figure}[t]
\centering
\includegraphics[width=0.46\textwidth]{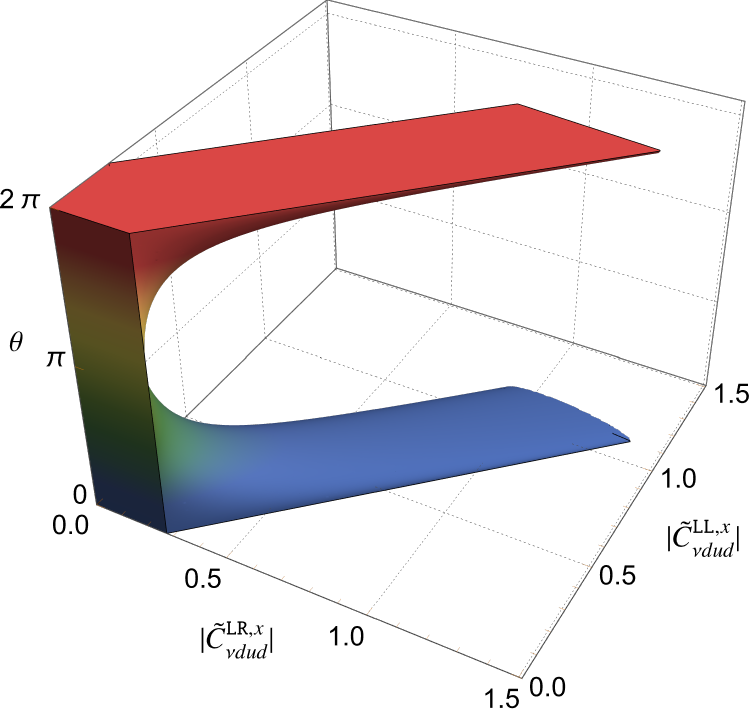}
\caption{The allowed region in the three-dimensional plot spanned by 
$\{|\tilde C^{\tL\tR,x}_{ \nu dud}|,\,|\tilde C^{\tL\tL,x}_{ \nu dud}|,\,\theta\}$, which is obtained by simultaneously satisfying the experimental bounds in \cref{eq:exp_bound_nu}.}
\label{fig:newbound3D}    
\end{figure}

For this purpose, we parametrize the two relevant WCs as $C^{\tL\tR,x}_{ \nu dud}\equiv|\tilde C^{\tL\tR,x}_{ \nu dud}| e^{i\theta_1}/(10^{15}\,{\rm GeV})^2$ and $C^{\tL\tL,x}_{ \nu dud}\equiv
|\tilde C^{\tL\tL,x}_{ \nu dud}| e^{i\theta_2}/(10^{15}\,{\rm GeV})^2$, with the phase difference $\theta\equiv\theta_1-\theta_2$, 
where $x=e,\,\mu,\,\tau$ represents the neutrino flavor and
$|\tilde C^{\tL\tR(\tL\tL),x}_{ \nu dud}|$ are real positive dimensionless parameters.
From \cref{eq:N2vbarM} in the Appendix, the decay widths for the 2 two-body modes in \cref{eq:exp_bound_nu} can be expressed as
\begin{subequations}
\begin{align}
\frac{\Gamma_{n \to \bar\nu_x \pi^0} }{10^{-65}\,\rm GeV} 
=&~ 19.7 \big|\tilde C^{\tL\tR,x}_{ \nu dud}\big|^2 
+ 20.1 \big|\tilde C^{\tL \tL,x}_{\nu dud}\big|^2 
\nonumber\\
& -39.8\,\cos\theta\, |\tilde C^{\tL \tR,x}_{\nu dud}|\,
|\tilde C^{\tL \tL,x}_{\nu dud}|,
\\
\frac{\Gamma_{n \to \bar\nu_x \eta} }{10^{-65}\,\rm GeV}
=&~0.095 \big|\tilde C^{\tL\tR,x}_{ \nu dud}\big|^2 
+ 8.38 \big|\tilde C^{\tL \tL,x}_{\nu dud}\big|^2 
\nonumber\\
&+1.79\,\cos\theta\, |\tilde C^{\tL \tR,x}_{\nu dud}|\,
|\tilde C^{\tL \tL,x}_{\nu dud}|.
\end{align}
\end{subequations}
We see that the three parameters $\{|\tilde C^{\tL\tR,x}_{ \nu dud}|,\,|\tilde C^{\tL\tL,x}_{ \nu dud}|,\,\theta\}$ completely determine the decay widths.   
Owing to the sign difference in the interference term between the two decay modes, the experimental lower bounds from \cref{eq:exp_bound_nu} confine the three-dimensional parameter space into a closed region, with the magnitudes of both WCs bounded from above.  
To visualize it, 
\cref{fig:newbound3D} shows the three-dimensional allowed region spanned by $\{|\tilde C^{\tL\tR,x}_{ \nu dud}|,\,|\tilde C^{\tL\tL,x}_{ \nu dud}|,\,\theta\}$, which clearly indicates that the magnitudes of both dimensionless WCs are less than about 1.5. 
By taking values for the three parameters within this allowed region, we calculate the partial lifetimes for the remaining 6 modes:
$p\to \bar\nu_x\pi^+$, 
$p\to \bar\nu_x\pi^+\pi^0$, 
$p\to \bar\nu_x\pi^+\eta$, 
$n\to \bar\nu_x\pi^+\pi^-$, 
$n\to \bar\nu_x\pi^0\pi^0$,
and $n\to \bar\nu_x\pi^0\eta$.
For each decay mode, the smallest partial lifetime obtained is taken as our final conservative bound.   

Note that we use these published 
90\,\% CL limits as hard inequalities and propagate them to obtain bounds on the correlated decay modes. These derived bounds should therefore not be interpreted as $90\,\%$ CL bounds. A rigorous $90\,\%$ CL construction would require a detailed analysis of the underlying experimental likelihoods, including the relevant statistical and systematic correlations and nuisance parameters.

\begin{figure}[t]
\centering
\includegraphics[width=0.48\textwidth]{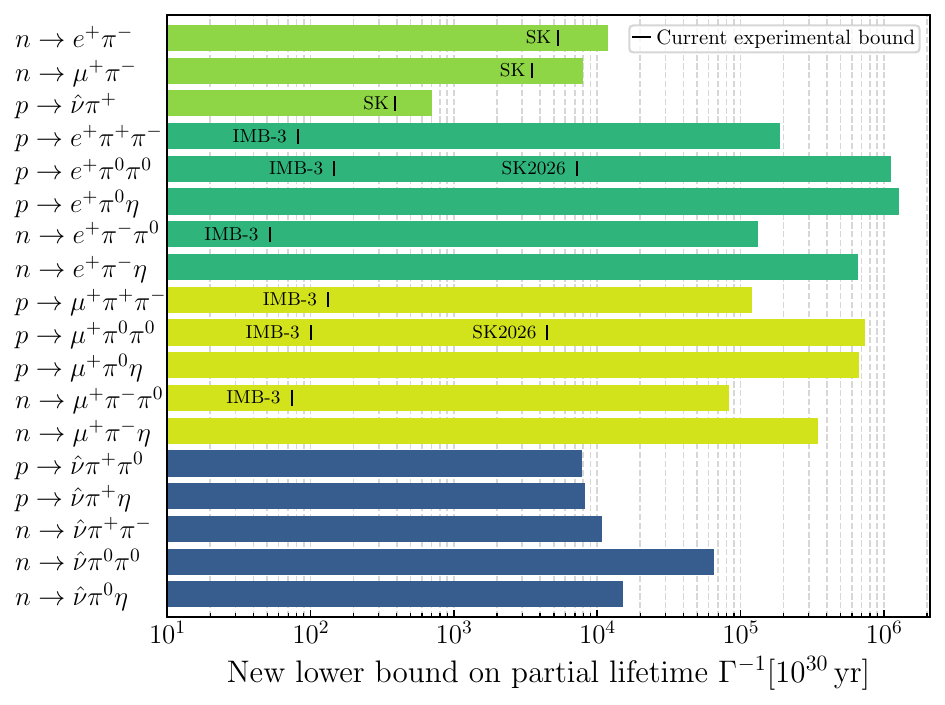}
\caption{The improved bounds on 18 two- and three-body nucleon decay modes based on the 6 best experimental constraints in \cref{eq:exp_bound_l,eq:exp_bound_nu}. The current 90\,\% CL experimental limits on two-body modes from Super-K (SK)~\cite{Super-Kamiokande:2017gev,Super-Kamiokande:2013rwg} and on three-body modes from IMB-3~\cite{McGrew:1999nd} are indicated by vertical solid lines. 
In addition, we include the latest Super-K 90\,\% CL bounds (SK2026) on the two three-body channels $p\to e^+\pi^0\pi^0$ and $p\to \mu^+\pi^0\pi^0$~\cite{Kamiokande:2026zwu}.
}
\label{fig:newbound3}    
\end{figure}

\textit{\textbf{Results}.} From the WC bounds derived above, we vary the WC values within the allowed regions to obtain conservative bounds on the partial lifetimes of the correlated decay modes. 
Our final improved bounds for the 15 three-body nucleon decay modes are summarized in \cref{fig:newbound3}. 
For the four charged-lepton modes
$p\to(e^+,\mu^+)\pi^+\pi^-$ and $n\to(e^+,\mu^+)\pi^-\pi^0$,
our indirect bounds are approximately 
$\Gamma^{-1}\gtrsim 10^{35}\,\rm yr$, 
which are more stringent than the previous IMB-3 bounds by roughly three orders of magnitude.
For $p\to (e^+,\mu^+)\pi^0\pi^0$, 
the bounds reach $10^{36}\,\rm yr$ and $7\times 10^{35}\,\rm yr$, respectively, exceeding the IMB-3 results by nearly four orders of magnitude. 
For the four charged-lepton modes involving an $\eta$ and the five neutrino/antineutrino modes, we provide stringent bounds for the first time.
Their partial lifetime bounds are 
$\Gamma^{-1}(p\to e^+\pi^0\eta)\gtrsim 10^{36}\,\rm yr$, 
$\Gamma^{-1}(n\to e^+\pi^-\eta)\gtrsim  6\times 10^{35}\,\rm yr$, 
$\Gamma^{-1}(p\to \mu^+\pi^0\eta)\gtrsim 6\times 10^{35}\,\rm yr$, 
$\Gamma^{-1}(n\to \mu^+\pi^-\eta)\gtrsim  3\times 10^{35}\,\rm yr$,  
$\Gamma^{-1}(n\to\hat\nu\pi^0\pi^0)\gtrsim 6\times 10^{34}\,\rm yr$, 
and $\Gamma^{-1} \gtrsim 10^{34}\,\rm yr$ for $p\to\hat\nu\pi^+(\pi^0,\eta)$, 
$n\to\hat\nu\pi^+\pi^-$, and
$n\to\hat\nu\pi^0\eta$.
If Super-K in future searches improves the sensitivity limit on $n\to \hat\nu\eta$ by an order of magnitude over the IMB result, our obtained bounds on the two $\eta$-related modes $p\to \hat\nu \pi^+ \eta $ and $n\to \hat\nu \pi^0\eta $ would also be strengthened by nearly an order of magnitude.
It is worth noting that the comparison in \cref{fig:newbound3} is intended to illustrate the numerical strength of our derived bounds relative to the existing experimental limits, rather than to compare bounds having identical statistical coverage.

Interestingly, 
although for any given WC the decay widths for modes involving an $\eta$ are suppressed by one to three orders of magnitude compared to $\pi$-only modes (see the Appendix), 
the derived conservative bounds are comparable. This is because, within the WC region allowed by the two-body modes, each three-body mode attains its maximal decay width at a distinct point in the multidimensional space.

In addition to the three-body modes, for the 3 two-body modes 
$n\to (e^+,\mu^+)\pi^-$ and $p\to \hat\nu\pi^+$, our global analysis leads to the improved bounds
$\Gamma^{-1}(n\to e^+\pi^-)\gtrsim 
1.2\times 10^{34}\,\rm yr$, 
$\Gamma^{-1}(n\to \mu^+\pi^-)\gtrsim
8\times 10^{33}\,\rm yr$,
and 
$\Gamma^{-1}(p\to \hat\nu\pi^+)\gtrsim 
7\times 10^{32}\,\rm yr$. 
They are approximately twice as strong as the current experimental bounds
$\Gamma^{-1}_{\tt SK}(n\to e^+\pi^-)\gtrsim 5.3\times 10^{33}\,\rm yr$, 
$\Gamma^{-1}_{\tt SK}(n\to \mu^+\pi^-)\gtrsim 3.5\times 10^{33}\,\rm yr$~\cite{Super-Kamiokande:2017gev},
and $\Gamma^{-1}_{\tt SK}(p\to \hat\nu\pi^+)\gtrsim 3.9\times 10^{32}\,\rm yr$~\cite{Super-Kamiokande:2013rwg}.
By isospin symmetry, these modes are respectively related to the proton and neutron decay modes in \cref{eq:exp_bound_l,eq:exp_bound_nu} via the relations: 
$\Gamma(n\to e^+ \pi^-)\approx 2\Gamma(p\to e^+ \pi^0)$,  
$\Gamma(n\to \mu^+ \pi^-)\approx 2\Gamma(p\to \mu^+ \pi^0)$, 
and $\Gamma(p\to \hat\nu \pi^+)\approx 2\Gamma(n\to \hat\nu \pi^0)$. 
As seen in \cref{eq:exp_bound_l,eq:exp_bound_nu}, our newly derived bounds perfectly satisfy these relations, which further strengthens our method and the established bounds on those three-body modes.  
Regarding the omitted vector-meson contributions and hadronic uncertainties,
we find that these effects alter the derived bounds by at most an $\calO(1)$ factor~\cite{Fan:2026csl}.

Before closing this part, we emphasize that our indirect bounds should be viewed as EFT-consistency limits. If a future observation of a three-body decay were to lie above its indirect bound, it would indicate failure of some assumptions underlying our analysis, such as dominance of dim-6 LEFT operators, leading-order chiral approximation, or absence of some exotic new-physics contributions.

\textit{\textbf{Conclusion}.} In this paper, we have presented a novel investigation of nucleon decays into one lepton and two nonstrange mesons. 
Using dim-6 BNV operators as our starting point, we established correlations between the well-constrained two-body and less-constrained three-body decays of the nucleons. 
By confining the multidimensional WC space to a closed, bounded region from the existing best bounds on two-body modes, we derived stringent lower bounds on partial lifetimes of the 15 three-body nucleon decay modes involving $\pi\pi$ or $\pi\eta$ final states.  
Our final results are summarized in \cref{fig:newbound3}.
In particular,  
the newly established indirect bounds on $p\to e^+\pi^0\pi^0$ and $p\to \mu^+\pi^0\pi^0$ are more stringent than the previous IMB-3 results (adopted by the Particle Data Group) by more than three orders of magnitude. 
Moreover, these limits are also stronger than the latest Super-K results by nearly two orders of magnitude~\cite{Kamiokande:2026zwu}.  

Our results set a benchmark for future experimental search of these decay modes. 
It is no exaggeration to say that these three-body 
channels are as important as the two-body modes. 
If the two-body modes are confirmed in future experiments, then these three-body modes, together with the two-body modes involving a vector meson~\cite{Liao:2025sqt}, can be used to help fully resolve operator degeneracies and determine the relevant operator structure responsible for nucleon decay.

\textit{\textbf{Note added}.} During the finalization of this manuscript,
a preprint by the Kamiokande Collaboration appeared~\cite{Kamiokande:2026zwu}, providing new experimental bounds on $p\to e^+\pi^0\pi^0$ and $p\to \mu^+\pi^0\pi^0$.
We have incorporated their bounds into our key results in \cref{fig:newbound3} for comparison. 

\textit{\textbf{Acknowledgments}} This work was supported by Grants No.\,NSFC-12305110 and No.\,NSFC-12035008.  

\textit{\textbf{Data availability}} There are no publicly available research data or software supporting this manuscript. Requests for further information or data should be sent to the authors.

\onecolumngrid
\begin{center}
\vspace{0.1cm}{\bf Appendix}
\end{center}

The complete expressions for the decay widths of all 24 decay modes, comprising 9 two-body modes and 15 three-body modes, are summarized below. 
\begin{itemize}
\item[(i)] {$e^+$ case}\\
Two-body modes:
\begin{subequations}
\begin{align}
\frac{\Gamma_{p \to e^+ \pi^0} }{(0.1\, \rm GeV)^5} 
=&~ 19.7 \big|C^{\tL\tR,e}_{\ell uud}\big|^2 + 
20.1 \big|C^{\tL \tL,e}_{\ell uud}\big|^2 
-39.8 \,\Re\big( C^{\tL \tR,e}_{\ell uud} C^{\tL \tL,e*}_{\ell uud} \big)
-0.0013 C^{\tL \tR,e}_{\ell uud} C^{\tR \tL,e*}_{\ell uud}
\label{eq:p2epi0}
\nonumber\\
& +0.0013 \big( C^{\tL \tR,e}_{\ell uud} C^{\tR \tR,e*}_{\ell uud} + 
C^{\tL \tL,e}_{\ell uud} C^{\tR \tL,e*}_{\ell uud} 
- C^{\tL \tL,e}_{\ell uud} C^{\tR \tR,e*}_{\ell uud} \big)
+\tL \leftrightarrow \tR,
\\
\frac{\Gamma_{p \to e^+ \eta} }{(0.1\, \rm GeV)^5} 
=&~0.095 \big|C^{\tL\tR,e}_{\ell uud}\big|^2 
+ 8.3 \big|C^{\tL \tL,e}_{\ell uud}\big|^2 
+1.8 \,\Re\big( C^{expressions\tL \tR,e}_{\ell uud} C^{\tL \tL,e*}_{\ell uud} \big)
+4.8 \cdot 10^{-4} C^{\tL \tR,e}_{\ell uud} C^{\tR \tL,e*}_{\ell uud}
\nonumber\\
& +0.0028 \big( C^{\tL \tR,e}_{\ell uud} C^{\tR \tR,e*}_{\ell uud} + 
C^{\tL \tL,e}_{\ell uud} C^{\tR \tL,e*}_{\ell uud} \big)
+ 0.011 C^{\tL \tL,e}_{\ell uud} C^{\tR \tR,e*}_{\ell uud}
+\tL \leftrightarrow \tR,
\\
\frac{\Gamma_{n \to e^+ \pi^-} }{(0.1\, \rm GeV)^5} 
=&~ 39.4 \big|C^{\tL\tR,e}_{\ell uud}\big|^2 
+ 40.2 \big|C^{\tL \tL,e}_{\ell uud}\big|^2 
-79.6 \,\Re\big( C^{\tL \tR,e}_{\ell uud} C^{\tL \tL,e*}_{\ell uud} \big)
-0.0026 C^{\tL \tR,e}_{\ell uud} C^{\tR \tL,e*}_{\ell uud}
\nonumber\\
& +0.0026 \big( C^{\tL \tR,e}_{\ell uud} C^{\tR \tR,e*}_{\ell uud} + 
C^{\tL \tL,e}_{\ell uud} C^{\tR \tL,e*}_{\ell uud} 
- C^{\tL \tL,e}_{\ell uud} C^{\tR \tR,e*}_{\ell uud} \big)
+\tL \leftrightarrow \tR.
\end{align}
\end{subequations}
Three-body modes:
\begin{subequations}
\begin{align}
\frac{\Gamma_{p \to e^+ \pi^+\pi^-} }{(0.1\, \rm GeV)^5} 
=&~ 2.52 \big|C^{\tL\tR,e}_{\ell uud}\big|^2 + 
2.57 \big|C^{\tL \tL,e}_{\ell uud}\big|^2 
-5.08\,\Re\big( C^{\tL \tR,e}_{\ell uud} C^{\tL \tL,e*}_{\ell uud} \big)
+ 0.002 C^{\tL \tR,e}_{\ell uud} C^{\tR \tL,e*}_{\ell uud}
\nonumber\\
& -0.0021 \big( C^{\tL \tR,e}_{\ell uud} C^{\tR \tR,e*}_{\ell uud} + 
C^{\tL \tL,e}_{\ell uud} C^{\tR \tL,e*}_{\ell uud} 
- C^{\tL \tL,e}_{\ell uud} C^{\tR \tR,e*}_{\ell uud} \big)
+\tL \leftrightarrow \tR,
\label{eq:p2epippim}
\\
\frac{\Gamma_{p \to e^+ \pi^0\pi^0} }{(0.1\, \rm GeV)^5} 
=&~ 0.42 \big|C^{\tL\tR,e}_{\ell uud}\big|^2 + 
0.43 \big|C^{\tL \tL,e}_{\ell uud}\big|^2 
-0.85 \,\Re\big( C^{\tL \tR,e}_{\ell uud} C^{\tL \tL,e*}_{\ell uud} \big)
+ 1.31\cdot 10^{-4} C^{\tL \tR,e}_{\ell uud} C^{\tR \tL,e*}_{\ell uud}
\nonumber\\
&-1.32\cdot 10^{-4} \big( C^{\tL \tR,e}_{\ell uud} C^{\tR \tR,e*}_{\ell uud} + 
C^{\tL \tL,e}_{\ell uud} C^{\tR \tL,e*}_{\ell uud}  \big)
+1.33\cdot 10^{-4} C^{\tL \tL,e}_{\ell uud} C^{\tR \tR,e*}_{\ell uud}
+\tL \leftrightarrow \tR,
\label{eq:p2epi0pi0}
\\
\frac{\Gamma_{p \to e^+ \pi^0\eta} }{(0.1\, \rm GeV)^5} 
=&~ 0.0043 \big|C^{\tL\tR,e}_{\ell uud}\big|^2 + 
0.064 \big|C^{\tL \tL,e}_{\ell uud}\big|^2 
+0.032 \,\Re\big( C^{\tL \tR,e}_{\ell uud} C^{\tL \tL,e*}_{\ell uud} \big)
-2.19 \cdot 10^{-6} C^{\tL \tR,e}_{\ell uud} C^{\tR \tL,e*}_{\ell uud}
\nonumber\\
& -7.86 \cdot 10^{-6} \big( C^{\tL \tR,e}_{\ell uud} C^{\tR \tR,e*}_{\ell uud} + C^{\tL \tL,e}_{\ell uud} C^{\tR \tL,e*}_{\ell uud} \big)
-2.04 \cdot 10^{-5} C^{\tL \tL,e}_{\ell uud} C^{\tR \tR,e*}_{\ell uud}
+\tL \leftrightarrow \tR,
\\
\frac{\Gamma_{n \to e^+ \pi^-\pi^0} }{(0.1\, \rm GeV)^5} 
=&~ 3.55 \big|C^{\tL\tR,e}_{\ell uud}\big|^2 
+ 3.61 \big|C^{\tL \tL,e}_{\ell uud}\big|^2 
-7.16 \,\Re\big( C^{\tL \tR,e}_{\ell uud} C^{\tL \tL,e*}_{\ell uud} \big)
+ 0.0037 C^{\tL \tR,e}_{\ell uud} C^{\tR \tL,e*}_{\ell uud}
\nonumber\\
& -0.0037 \big( C^{\tL \tR,e}_{\ell uud} C^{\tR \tR,e*}_{\ell uud} + 
C^{\tL \tL,e}_{\ell uud} C^{\tR \tL,e*}_{\ell uud} 
- C^{\tL \tL,e}_{\ell uud} C^{\tR \tR,e*}_{\ell uud} \big)
+\tL \leftrightarrow \tR,
\\
\frac{\Gamma_{n \to e^+ \pi^-\eta} }{(0.1\, \rm GeV)^5} 
=&~ 0.0083 \big|C^{\tL\tR,e}_{\ell uud}\big|^2 
+ 0.124 \big|C^{\tL \tL,e}_{\ell uud}\big|^2 
+ 0.062 \,\Re\big( C^{\tL \tR,e}_{\ell uud} C^{\tL \tL,e*}_{\ell uud} \big)
-4.57 \cdot 10^{-6} C^{\tL \tR,e}_{\ell uud} C^{\tR \tL,e*}_{\ell uud}
\nonumber\\
& -1.62 \cdot 10^{-5} \big( C^{\tL \tR,e}_{\ell uud} C^{\tR \tR,e*}_{\ell uud} + C^{\tL \tL,e}_{\ell uud} C^{\tR \tL,e*}_{\ell uud} \big)
-4.22 \cdot 10^{-5} C^{\tL \tL,e}_{\ell uud} C^{\tR \tR,e*}_{\ell uud}
+\tL \leftrightarrow \tR.
\end{align}
\end{subequations}

\item[(ii)] {$\mu^+$ case}\\
Two-body modes:
\begin{subequations}
\begin{align}
\frac{\Gamma_{p \to \mu^+ \pi^0} }{(0.1\, \rm GeV)^5} 
=&~19.4\big|C^{\tL\tR,\mu}_{\ell uud}\big|^2 
+ 19.8\big|C^{\tL \tL,\mu}_{\ell uud}\big|^2 
-39.2\,\Re\big( C^{\tL \tR,\mu}_{\ell uud} C^{\tL \tL,\mu*}_{\ell uud} \big)
-0.26 C^{\tL \tR,\mu}_{\ell uud} C^{\tR \tL,\mu*}_{\ell uud}
\nonumber\\
& +0.26 \big( C^{\tL \tR,\mu}_{\ell uud} C^{\tR \tR,\mu*}_{\ell uud} + C^{\tL \tL,\mu}_{\ell uud} C^{\tR \tL,\mu*}_{\ell uud} \big)
- 0.27 C^{\tL \tL,\mu}_{\ell uud} C^{\tR \tR,\mu*}_{\ell uud}
+ \tL \leftrightarrow \tR,
\\
\frac{\Gamma_{p \to \mu^+ \eta} }{(0.1\, \rm GeV)^5} 
=&~ 0.12 \big|C^{\tL\tR,\mu}_{\ell uud}\big|^2 
+ 8.03 \big|C^{\tL \tL,\mu}_{\ell uud}\big|^2 
+ 1.8 \,\Re\big( C^{\tL \tR,\mu}_{\ell uud} C^{\tL \tL,\mu*}_{\ell uud} \big)
+0.097 C^{\tL \tR,\mu}_{\ell uud} C^{\tR \tL,\mu*}_{\ell uud}
\nonumber\\
& +0.56 \big( C^{\tL \tR,\mu}_{\ell uud} C^{\tR \tR,\mu*}_{\ell uud} + C^{\tL \tL,\mu}_{\ell uud} C^{\tR \tL,\mu*}_{\ell uud} \big)
+2.1 C^{\tL \tL,\mu}_{\ell uud} C^{\tR \tR,\mu*}_{\ell uud}
+ \tL \leftrightarrow \tR,
\\
\frac{\Gamma_{n \to \mu^+ \pi^-} }{(0.1\, \rm GeV)^5} 
=&~ 38.85 \big|C^{\tL\tR,\mu}_{\ell uud}\big|^2 
+ 39.6 \big|C^{\tL \tL,\mu}_{\ell uud}\big|^2 
-78.4 \,\Re\big( C^{\tL \tR,\mu}_{\ell uud} C^{\tL \tL,\mu*}_{\ell uud} \big)
-0.52 C^{\tL \tR,\mu}_{\ell uud} C^{\tR \tL,\mu*}_{\ell uud}
\nonumber\\
& +0.53 \big( C^{\tL \tR,\mu}_{\ell uud} C^{\tR \tR,\mu*}_{\ell uud} + C^{\tL \tL,\mu}_{\ell uud} C^{\tR \tL,\mu*}_{\ell uud} 
- C^{\tL \tL,\mu}_{\ell uud} C^{\tR \tR,\mu*}_{\ell uud} \big)
+ \tL \leftrightarrow \tR.
\end{align}
\end{subequations}
Three-body modes:
\begin{subequations}
\begin{align}
\frac{\Gamma_{p \to \mu^+ \pi^+\pi^-} }{(0.1\, \rm GeV)^5} 
=&~ 2.24 \big|C^{\tL\tR,\mu}_{\ell uud}\big|^2 + 
2.29 \big|C^{\tL \tL,\mu}_{\ell uud}\big|^2 
-4.53\,\Re\big( C^{\tL \tR,\mu}_{\ell uud} C^{\tL \tL,\mu*}_{\ell uud} \big)
+ 0.318 C^{\tL \tR,\mu}_{\ell uud} C^{\tR \tL,\mu*}_{\ell uud}
\nonumber\\
& -0.321 \big( C^{\tL \tR,\mu}_{\ell uud} C^{\tR \tR,\mu*}_{\ell uud} + C^{\tL \tL,\mu}_{\ell uud} C^{\tR \tL,\mu*}_{\ell uud} \big)
+ 0.324 C^{\tL \tL,\mu}_{\ell uud} C^{\tR \tR,\mu*}_{\ell uud}
+ \tL \leftrightarrow \tR,
\\
\frac{\Gamma_{p \to \mu^+ \pi^0\pi^0} }{(0.1\, \rm GeV)^5} 
=&~ 0.39 \big|C^{\tL\tR,\mu}_{\ell uud}\big|^2 + 
0.4 \big|C^{\tL \tL,\mu}_{\ell uud}\big|^2 
-0.79 \,\Re\big( C^{\tL \tR,\mu}_{\ell uud} C^{\tL \tL,\mu*}_{\ell uud} \big)
+ 0.024 C^{\tL \tR,\mu}_{\ell uud} C^{\tR \tL,\mu*}_{\ell uud}
\nonumber\\
& -0.024 \big( C^{\tL \tR,\mu}_{\ell uud} C^{\tR \tR,\mu*}_{\ell uud} + C^{\tL \tL,\mu}_{\ell uud} C^{\tR \tL,\mu*}_{\ell uud} 
-C^{\tL \tL,\mu}_{\ell uud} C^{\tR \tR,\mu*}_{\ell uud} \big)
+ \tL \leftrightarrow \tR,
\\
\frac{\Gamma_{p \to \mu^+ \pi^0\eta} }{(0.1\, \rm GeV)^5} 
=&~ 0.0027 \big|C^{\tL\tR,\mu}_{\ell uud}\big|^2 + 
0.039 \big|C^{\tL \tL,\mu}_{\ell uud}\big|^2 
+ 0.02 \,\Re\big( C^{\tL \tR,\mu}_{\ell uud} C^{\tL \tL,\mu*}_{\ell uud} \big)
-3.52 \cdot 10^{-4} C^{\tL \tR,\mu}_{\ell uud} C^{\tR \tL,\mu*}_{\ell uud}
\nonumber\\
& -0.0013 \big( C^{\tL \tR,\mu}_{\ell uud} C^{\tR \tR,\mu*}_{\ell uud} + C^{\tL \tL,\mu}_{\ell uud} C^{\tR \tL,\mu*}_{\ell uud} \big)
-0.0039 C^{\tL \tL,\mu}_{\ell uud} C^{\tR \tR,\mu*}_{\ell uud}
+ \tL \leftrightarrow \tR,
\\
\frac{\Gamma_{n \to \mu^+ \pi^-\pi^0} }{(0.1\, \rm GeV)^5} 
=&~ 3.12 \big|C^{\tL\tR,\mu}_{\ell uud}\big|^2 
+ 3.18 \big|C^{\tL \tL,\mu}_{\ell uud}\big|^2 
-6.29\,\Re\big( C^{\tL \tR,\mu}_{\ell uud} C^{\tL \tL,\mu*}_{\ell uud} \big)
+ 0.559 C^{\tL \tR,\mu}_{\ell uud} C^{\tR \tL,\mu*}_{\ell uud}
\nonumber\\
& -0.564 \big( C^{\tL \tR,\mu}_{\ell uud} C^{\tR \tR,\mu*}_{\ell uud} + C^{\tL \tL,\mu}_{\ell uud} C^{\tR \tL,\mu*}_{\ell uud} \big)
+ 0.569 C^{\tL \tL,\mu}_{\ell uud} C^{\tR \tR,\mu*}_{\ell uud}
+ \tL \leftrightarrow \tR,
\\
\frac{\Gamma_{n \to \mu^+ \pi^-\eta} }{(0.1\, \rm GeV)^5} 
=&~ 0.0052 \big|C^{\tL\tR,\mu}_{\ell uud}\big|^2 
+ 0.074 \big|C^{\tL \tL,\mu}_{\ell uud}\big|^2 
+ 0.038\,\Re\big( C^{\tL \tR,\mu}_{\ell uud} C^{\tL \tL,\mu*}_{\ell uud} \big)
-7.15 \cdot 10^{-4} C^{\tL \tR,\mu}_{\ell uud} C^{\tR \tL,\mu*}_{\ell uud}
\nonumber\\
& -0.0026 \big( C^{\tL \tR,\mu}_{\ell uud} C^{\tR \tR,\mu*}_{\ell uud} + C^{\tL \tL,\mu}_{\ell uud} C^{\tR \tL,\mu*}_{\ell uud} \big)
-0.0079 C^{\tL \tL,\mu}_{\ell uud} C^{\tR \tR,\mu*}_{\ell uud}
+ \tL \leftrightarrow \tR.
\end{align}
\end{subequations}
\item[(iii)] {$\bar\nu$ case}\\
Two-body modes:
\begin{subequations}
\label{eq:N2vbarM}
\begin{align}
\frac{\Gamma_{p \to \bar\nu_x \pi^+} }{(0.1\, \rm GeV)^5} 
=&~ 39.2 \big|C^{\tL\tR,x}_{ \nu dud}\big|^2 
+ 40 \big|C^{\tL \tL,x}_{\nu dud}\big|^2 
-79.2\,\Re\big( C^{\tL \tR,x}_{\nu dud} C^{\tL \tL,x*}_{\nu dud} \big),
\\
\frac{\Gamma_{n \to \bar\nu_x \pi^0} }{(0.1\, \rm GeV)^5} 
=&~ 19.7 \big|C^{\tL\tR,x}_{ \nu dud}\big|^2 
+ 20.1 \big|C^{\tL \tL,x}_{\nu dud}\big|^2 
-39.8\,\Re\big( C^{\tL \tR,x}_{\nu dud} C^{\tL \tL,x*}_{\nu dud} \big),
\\
\frac{\Gamma_{n \to \bar\nu_x \eta} }{(0.1\, \rm GeV)^5} 
=&~0.095 \big|C^{\tL\tR,x}_{ \nu dud}\big|^2 
+ 8.38 \big|C^{\tL \tL,x}_{\nu dud}\big|^2 
+1.79\,\Re\big( C^{\tL \tR,x}_{\nu dud} C^{\tL \tL,x*}_{\nu dud} \big).
\end{align}
\end{subequations}
Three-body modes: 
\begin{subequations}
\begin{align}
\frac{\Gamma_{p \to \bar\nu_x \pi^+\pi^0} }{(0.1\, \rm GeV)^5} 
=&~ 3.49 \big|C^{\tL\tR,x}_{ \nu dud}\big|^2 
+ 3.56\big|C^{\tL \tL,x}_{\nu dud}\big|^2 
-7.06\,\Re\big( C^{\tL \tR,x}_{\nu dud} C^{\tL \tL,x*}_{\nu dud} \big) ,
\\
\frac{\Gamma_{p \to \bar\nu_x \pi^+\eta} }{(0.1\, \rm GeV)^5}
=&~0.008 \big|C^{\tL \tR,x}_{\nu dud}\big|^2 
+ 0.12\big|C^{\tL \tL,x}_{\nu dud}\big|^2 
+0.06\,\Re\big( C^{\tL \tR,x}_{\nu dud} C^{\tL \tL,x*}_{\nu dud} \big),
\\
\frac{\Gamma_{n \to \bar\nu_x \pi^+\pi^-} }{(0.1\, \rm GeV)^5}
=&~2.56\big|C^{\tL \tR,x}_{\nu dud}\big|^2
+ 2.61\big|C^{\tL \tL,x}_{\nu dud}\big|^2 
- 5.17\,\Re\big( C^{\tL \tR,x}_{\nu dud} C^{\tL \tL,x*}_{\nu dud} \big),
\\
\frac{\Gamma_{n \to \bar\nu_x \pi^0\pi^0} }{(0.1\, \rm GeV)^5}
=&~ 0.42\big|C^{\tL \tR,x}_{\nu dud}\big|^2
+ 0.43\big|C^{\tL \tL,x}_{\nu dud}\big|^2 
- 0.86 \,\Re\big( C^{\tL \tR,x}_{\nu dud} C^{\tL \tL,x*}_{\nu dud} \big),
\\
\frac{\Gamma_{n \to \bar\nu_x \pi^0\eta} }{(0.1\, \rm GeV)^5}
=&~ 0.0043\big|C^{\tL \tR,x}_{\nu dud}\big|^2
+ 0.065\big|C^{\tL \tL,x}_{\nu dud}\big|^2 
+ 0.033 \,\Re\big( C^{\tL \tR,x}_{\nu dud} C^{\tL \tL,x*}_{\nu dud} \big).
\end{align}
\end{subequations}
The results for the neutrino modes are identical under the simultaneous interchange of $\nu\leftrightarrow \bar\nu$ and $\tL\leftrightarrow \tR$.
\end{itemize}

The decay widths for these $\Delta(B-L)=0$ two-body decays were also provided in ~\cite{Nath:2006ut}, where the lepton masses were neglected. 
Upon accounting for differences in operator conventions and adopting the same massless lepton approximation, we confirm agreement between our results and those in~\cite{Nath:2006ut}, with the exception of the $n\to (e^+,\mu^+) \pi^-$ channels. For these two modes, we notice a redundant factor of $3/2$ in their Eq.\,(538).

\vspace{1cm}
\twocolumngrid
\bibliography{reference}{}
\bibliographystyle{utphys}
\end{document}